# Boron Liquid Metal Alloy Ion Sources For Special FIB Applications

L. Bischoff [a)], N. Klingner,

Helmholtz-Zentrum Dresden-Rossendorf, Institute of Ion Beam Physics and Materials Research, Bautzner Landstrasse 400, 01328 Dresden, Germany

P. Mazarov [b)], W. Pilz, F. Meyer

Raith GmbH, Konrad-Adenauer-Allee 8, 44263 Dortmund, Germany

[a)] Electronic mail: l.bischoff@hzdr.de
[b)] Electronic mail: Paul.Mazarov@raith.de

Focused Ion Beam (FIB) processing has been established as a well-suited and promising technique in R&D in nearly all fields of nanotechnology for patterning and prototyping on the micro and nanometer scale and below. Among other concepts, liquid metal alloy ion sources (LMAIS) are one of the alternatives to conventional gallium beams to extend the FIB application field. To meet the rising demand for light ions, different boron containing alloys were investigated in this work. A promising solution was found in a $Co_{31}Nd_{64}B_5$ based LMAIS which will be introduced in more detail. Besides cobalt as a ferromagnetic element and the rare earth element neodymium, boron in particular is of interest for special FIB applications like local p-type doping, for which resolution of about 30 nm has been achieved so far.

## I. INTRODUCTION

Focused Ion Beam (FIB) devices are well-established in nanotechnology for nanostructuring, local surface modification, doping, prototyping as well as for ion beam analysis. The main component of such a FIB system is the ion source and the available ion species it can provide [1]. Most of the instruments currently work with a gallium liquid metal ion source (Ga-LMIS), but the demand for other ion species is increasing [2]. A particular element of interest is boron, one of the lightest elements in the periodic table and well established in microelectronics for p-type doping in silicon by implantation or diffusion [3]. There is a long term interest and many efforts were made for the application of boron in a LMAIS for local material modification by FIB in order to avoid boron broad beam implantation and the related lithography steps. Boron has two stable isotopes with a mass of 10 u (19.9% natural abundance) and 11 u (80.1% natural abundance). The melting point of boron is $T_{melt}$ = 2080°C with a vapor pressure of about $10^{-8}$ mbar at 1282°C. For operation in an ion source a considerably lower temperature is needed which can be obtained by suitable alloying. Palladium containing alloys like $B_{10}As_{10}Ni_{40}Pd_{40}$ and $B_{20}Ni_{40}Pd_{40}$ with a melting point of $T_{melt} = 650°C$ [4], $Ni_{45}B_{45}Si_{10}$ with $T_{melt} = 950°C$ [5] as well as the platinum alloy $Pt_{72}B_{28}$ with $T_{melt} = 790°C$ [6, 9] were investigated, tested and reported in the given references. A more complete list of boron LMAIS systems with corresponding references can be found in Ref. [2]. Unfortunately, only limited details on the operation, lifetime, emission currents or instabilities are published in these references. Finally, today there is no such boron containing ion source available on the market for FIB applications.

In this contribution various boron containing LMAIS candidates, which are of particular interest, were prepared, tested and characterized.

## II. EXPERIMENTAL

Hairpin emitters with different reservoir geometries were fabricated from 250 µm metallic wires and spot welded on a source heating filament. The tip was sharpened electrochemically for tungsten and rhenium material emitters and mechanically for tantalum based emitters to a final radius of $(4 \pm 1)$ µm. After heating to 1200°C in UHV ($< 10^{-7}$ mbar) the emitters were wetted in a crucible with the chosen alloys (see table 1). Slow heating of the crucible as well as of the emitter ($< 50$ K / minute) is necessary to obtain a homogeneous temperature distribution in the source reservoir. The filament and tip material was chosen with respect to the wetting behavior and chemical compatibility. In the same setup the emission was first tested against the cold crucible to minimize the contamination by back-sputtered material. The handling of the emitter for introducing it into a cartridge etc. has to be done at ambient pressure very quickly or in argon atmosphere. This includes the transportation of boron containing LMAIS in argon filled containers due to cleanliness requirements.

| Source material | $T_{melt}$ (°C) | Emitter | Content of Boron | |
|---|---|---|---|---|
| $Au_{77}Si_{18}B_5$ | 370 | Ta, W | $^{11}B^+/Au^+$ | $< 10^{-5}$ |
| $Au_{70}Ge_{25}B_5$ | 370 | Ta, W | $^{11}B^+/Au^+$ | $6 \times 10^{-4}$ |
| $Au_{68}Ge_{22}Ni_5B_5$ | 370 | Re | $^{11}B^+/Au^+$ | $4 \times 10^{-3}$ |
| $Co_{31}Nd_{64}B_5$ | 650 | Ta, Re, W | $^{11}B^+/Co^+$ | 0.1 |
| $Ni_{40}B_{60}$ | 1032 | W | No B emission | |

TABLE 1: Source materials, melting temperature, emitter tip material and content of boron intensity related to the main peak in the mass spectrum.

All loaded and wetted emitters had a good emission performance and were then introduced in a cartridge to characterize them in an analytic test-FIB described in [7]. Measurements of the *I-V* curve, the mass spectrum, current stability, lifetime and the full width at half maximum of the energy distribution for selected ion species depending on the emission current were carried out. The ratio of boron with respect to the main peak intensity is presented in table 1. The most promising candidate, the $Co_{31}Nd_{64}B_5$ LMAIS, was finally installed and investigated in detail in a mass-separated FIB/SEM system (VELION Raith GmbH) [8] where it is possible to switch between certain ion species by changing the parameters of the ExB mass filter.

## II. RESULTS AND DISCUSSION

### A. $Au_{77}Si_{18}B_5$ and $Au_{70}Ge_{25}B_5$

As a first attempt boron was added to classical metallic glasses (AuSi, AuGe) and tested on different emitter materials. Due to the low content of boron in the alloy the sources could be operated at temperatures slightly above the eutectic temperature of 365°C [9]. This low temperature is favorable due to the reduced out-diffusion and evaporation of boron. The relative intensity of the boron peaks is about 3 orders of magnitude smaller than gold, shown in Fig. 1 for AuGeB. In the case of AuSiB the relation was even worse so that both of these sources can not be used for practical applications.

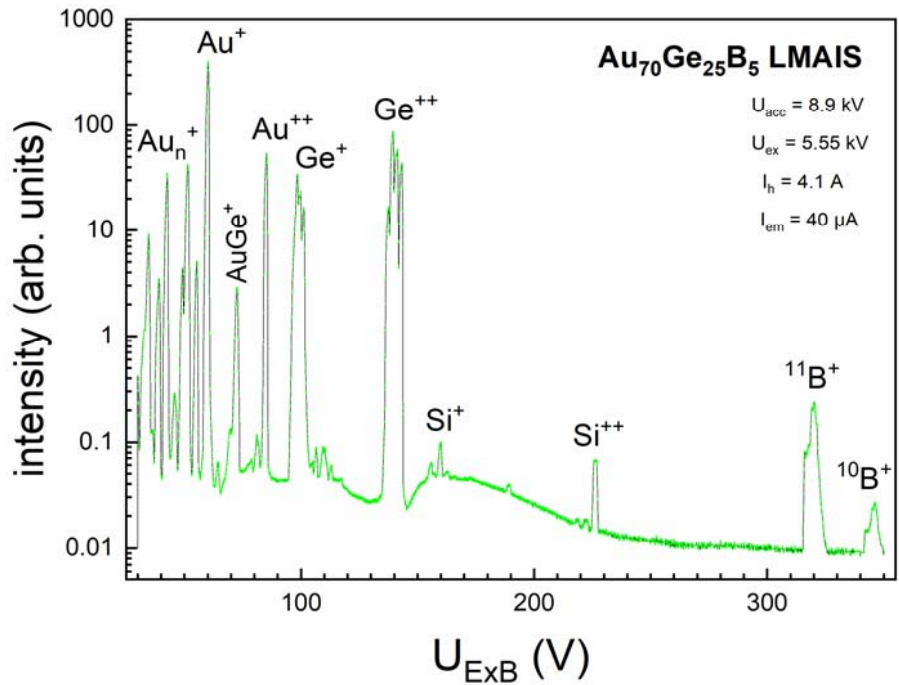

Figure 1: Mass spectrum of an $Au_{70}Ge_{25}B_5$ LMAIS. The current $I_{SEV}$ in the test-FIB as described in Ref. [7] was measured with a secondary electron multiplier and is given in arbitrary units. Here $U_{ExB}$ is the voltage on the ExB mass filter, $U_{acc}$ the acceleration voltage of the FIB, $U_{ex}$ the extraction voltage, $I_h$ the heating current of the LMAIS and $I_{em}$ the emission current.

## B. *Au$_{68}$Ge$_{22}$Ni$_5$B$_5$*

In a second approach a small amount of nickel was added to AuGeB to form an Au$_{68}$Ge$_{22}$Ni$_5$B$_5$ alloy. From previous investigations it is known that boron and nickel can work in a Ni$_{40}$B$_{60}$ LMAIS but it was used on a carbon tip at about 1030°C [10]. This new composition was chosen to lower the operation temperature and also to use a standard hair pin emitter with a metallic tip, rhenium in our case. A mass spectrum, obtained in the test-FIB [7] is shown in Fig. 2. The main operation parameters are given in the legend. The large number of lines in the spectrum due to isotopes and compound ions makes high resolution FIB applications difficult.

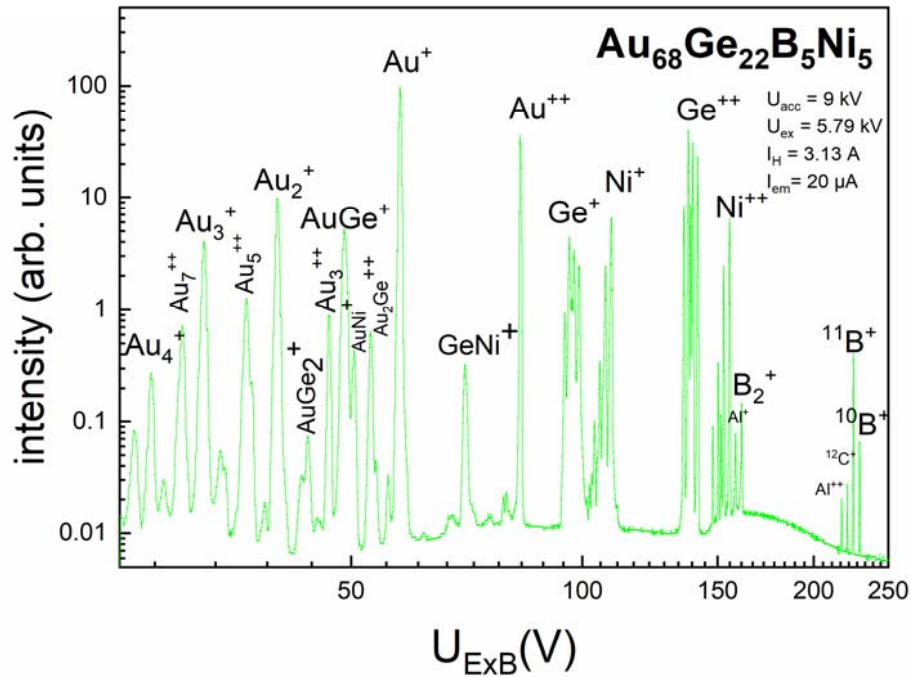

Figure 2: Mass spectrum of an Au$_{68}$Ge$_{22}$B$_5$ Ni$_5$ LMAIS obtained in the test FIB.

The operation of the source is limited to a temperature window between 400°C and 760°C. With lowering the heating power the total emission is strongly decreasing. At too high temperatures source material will be evaporated and NiGe precipitates as well [9]. In addition a higher emission current (20 µA) is required for a stable operation - unfavorable due to the resulting increase in chromatic aberrations. The total lifetime of the source was limited to about 10 hours. After long operation, restarting after an interruption and the formation of a new Taylor cone becomes increasingly difficult due to the changed composition of the alloy film at the emitter tip.

## C. Co$_{31}$Nd$_{64}$B$_5$

The most suitable source properties among the investigated candidates were found using a tungsten emitter wetted with a Co$_{31}$Nd$_{64}$B$_5$ alloy. The initial motivation for this source development was that the Co$_{36}$Nd$_{64}$ LMAIS ($T_{melt}$ = 566°C) [11] is very stable and reliable. Also this source needs some temperature restrictions to not exceed 700°C to prevent out diffusion and evaporation of boron. Furthermore, stable NdB will form which remains solid even above 1000°C. On the other hand a minimum temperature of 600°C is required to liquify the alloy. The emission current should be limited to 30 µA to avoid additional heating of the tip by for example secondary electron bombardment. The I-V characteristics of the Co$_{31}$Nd$_{64}$B$_5$ LMAIS obtained in the VELION system are shown in Fig. 3a. The slope of 0.143 µA/V is steeper than known from other sources. The reason is probably the large distance between tip and extractor of 3 mm resulting in high extraction voltages. A comparable value of 0.1 µA/V was also reported for a Pd$_{70}$As$_{16}$B$_{14}$ source at extraction potentials between 8.6 and 9 kV [12] which is explained with the presence of arsenic in the source material. The target current of $^{11}$B as a function of the heating current is presented in Fig. 3b. The emission starts at about 650° C, and at approximately 1000° C the emission saturates because mainly the alloy starts to evaporate. A working temperature higher than 700°C leads to a strong reduction of the source lifetime.

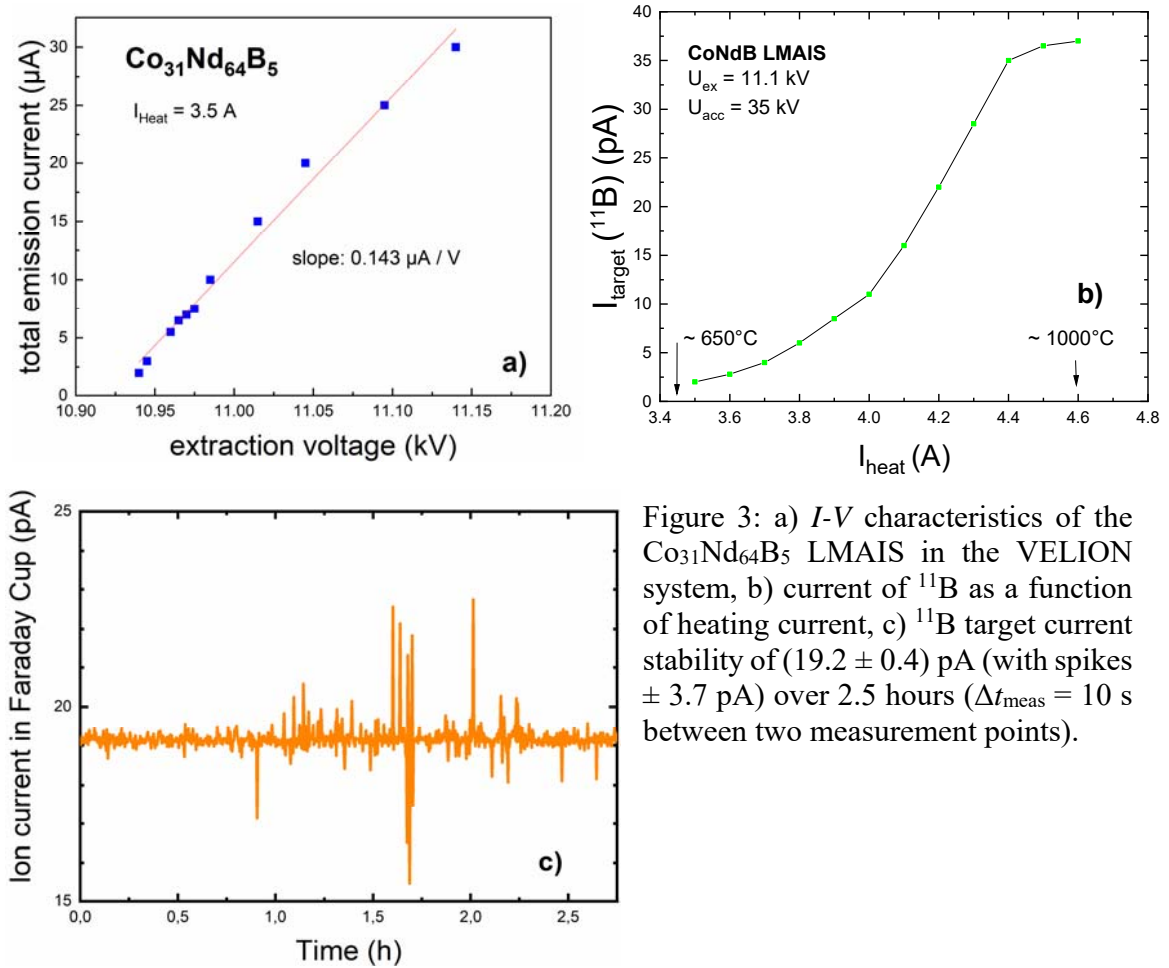

Figure 3: a) *I-V* characteristics of the Co$_{31}$Nd$_{64}$B$_5$ LMAIS in the VELION system, b) current of $^{11}$B as a function of heating current, c) $^{11}$B target current stability of (19.2 ± 0.4) pA (with spikes ± 3.7 pA) over 2.5 hours ($\Delta t_{meas}$ = 10 s between two measurement points).

The target current stability of the CoNdB LMAIS for the $^{11}$B beam was measured every 10 s over a time of 2.5 hours. It was determined to be $(19.2 \pm 0.4)$ pA apart from some spikes with a maximum deviation of $\pm 3.7$ pA, see Fig. 3c. The source could be operated in a stable mode over a duration of more than 600 µAh at an emission current of about 2 µA giving a total lifetime of more than 300 h.

The mass spectrum of the ion source was measured in our test-FIB system [7] and is presented in Fig. 4. The applied operation parameters are plotted in the inset. The $^{11}$B intensity reaches more than 5% of Nd$^{++}$ and 10% of Co$^{+}$ or Co$^{++}$. Due to contamination of the starting materials, a few additional lines in the percentage intensity range or less can be seen in high-resolution mass spectroscopy. Applying this source in the VELION system also the doubly charged boron species could be found shown in Fig. 5. The current in this FIB system is measured in a Faraday cup, i.e. the current for double charged ions must be divided by two to get the real number of particles per second. The maximum total ion current without mass separation was 4 nA at 10 µA emission current in the VELION system. In the case of the light ion boron the ionization energy from single to double charge state increases only from 8.3 eV to 25.1 eV [13], whereas in previous experiments for a Ga$_{35}$Bi$_{60}$Li$_5$ LMAIS [14] this increase for lithium was from 5.4 eV to 75.6 eV [13] whereby doubly charged lithium ions could not be detected.

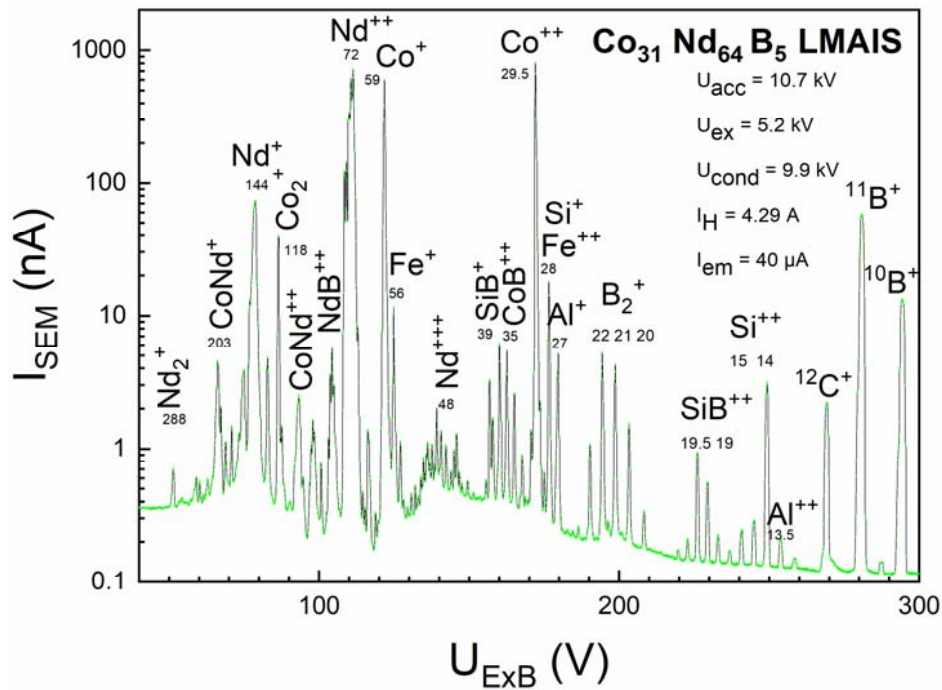

Figure 4: Mass spectrum of a Co$_{31}$Nd$_{64}$B$_5$ LMAIS. The current $I_{SEM}$ in the test-FIB was measured with a secondary electron multiplier (SEM).

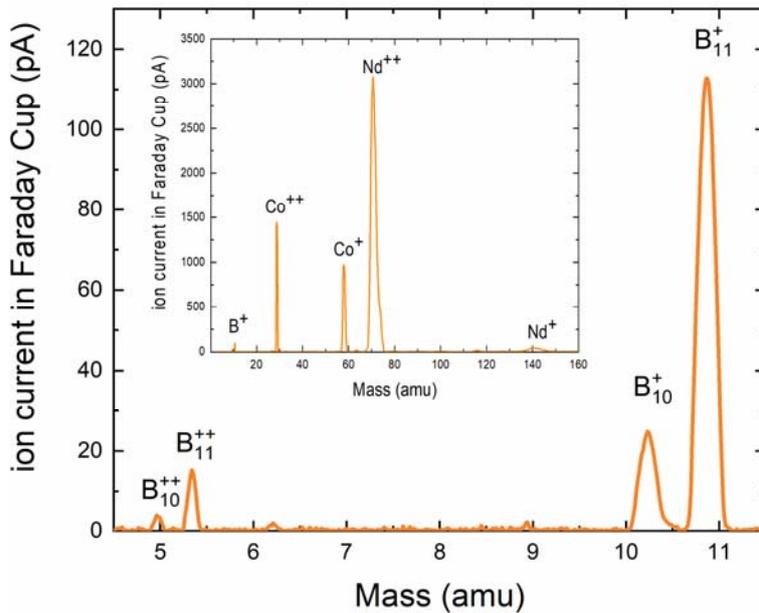

Figure 5: Mass spectrum of a Co$_{31}$Nd$_{64}$B$_5$ LMAIS obtained in the VELION system. The full spectrum is shown in the inset as well as the evidence of the single and double charged well separated boron isotopes.

Finally the FIB imaging resolution for the $^{11}$B$^+$ beam was estimated by imaging a Chessy-test structure (Plano) depicted in Fig. 6. For an emission current of 10 μA at an energy of 35 keV in the VELION system, the edge resolution (25% – 75% raise of intensity) was evaluated to (30 ± 5) nm for an $^{11}$B$^+$ FIB with 5 pA ion current, see Fig. 6a. The application of a Co$_{31}$Nd$_{64}$B$_5$ LMAIS gives the opportunity to use beside boron also heavier elements in the FIB. While these heavy elements lend themselves to milling applications they can also be applied for imaging. As an example in Fig. 6b the imaging of a Chessy-test structure using a $^{142}$Nd$^{++}$ FIB with 44 pA is presented.

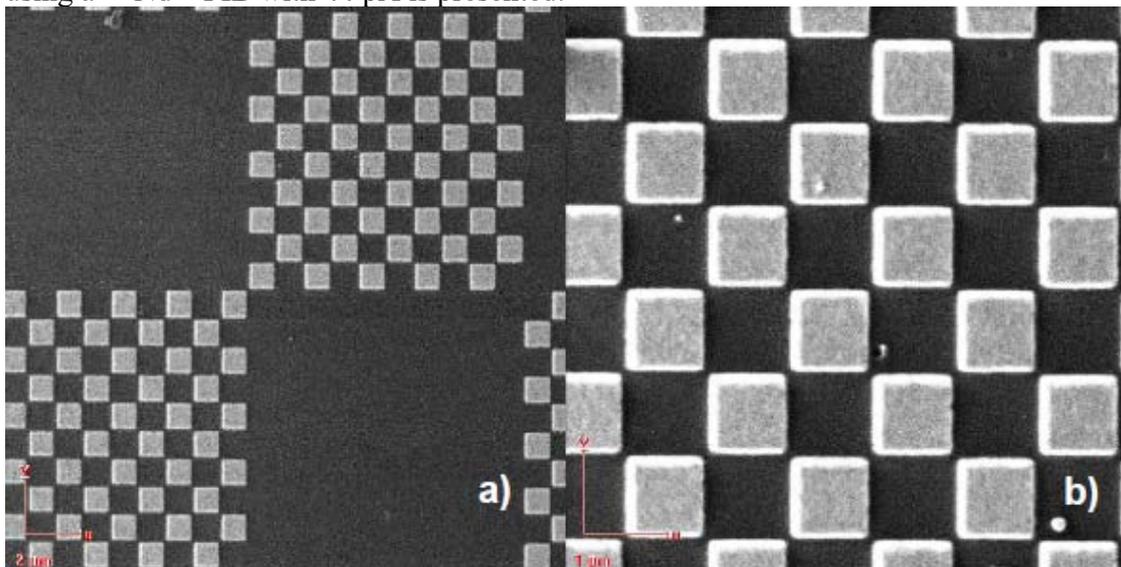

Fig. 6. a) Imaging of a Chessy-test structure with an $^{11}$B$^+$ FIB with 5 pA. b) Imaging of a Chessy-test structure with 44 pA of $^{142}$Nd$^{++}$.

## III. SUMMARY AND CONCLUSIONS

For the primary aim of achieving local p-type doping at the nanoscale, boron- containing alloys were tested in order to fabricate a long lasting and stable boron delivering LMAIS. Some possible candidates namely $Co_{31}Nd_{64}B_5$, $Au_{77}Si_{18}B_5$, $Au_{70}Ge_{25}B_5$ and $Au_{68}Ge_{22}Ni_5B_5$ were investigated. The first one exhibited the best performance and was additionally tested under realistic conditions in a VELION FIB-SEM system (Raith GmbH). The source lifetime was more than 600 µAh and a first characterization showed a lateral resolution of (30 ± 5) nm so far. For different FIB applications fast switching between the different ion species is possible to select the best candidate for the respective purpose. In the case of a $Co_{31}Nd_{64}B_5$ LMAIS, boron – very light is favorable for imaging, ion lithography [15] and local p-type doping [3]. Cobalt with medium mass can be used for applications in the field of nano-magnetics or $CoSi_2$ ion beam synthesis of conductive nano-structures on silicon [16]. Finally neodymium as single or double charged ion is the best choice for high rate sputtering and surface patterning because of its high mass. However, it is problematic to separate all neodymium isotopes leading to a reduced lateral resolution.

The change between ion species can be done in seconds and leads to remarkable expansion of the application spectrum of FIB technology. The application perspectives of such an ion source, like local boron doping by FIB will be examined in a prospective investigation.

## ACKNOWLEDGMENTS

The authors would like to thank the German Federal Ministry of Economics BMWi for financial support under grant no. ZF4330902DF7. Support by the Ion Beam Center, HZDR, is gratefully acknowledged.

This article has been submitted to the Journal of Vacuum Science & Technology B. After it is published, it will be found at: https://avs.scitation.org/journal/jvb

References:

[1] J. Gierak, Semicond. Sci. Tech. **24**, 043001 (2009).

[2] L. Bischoff, P. Mazarov, L. Bruchhaus, and J. Gierak, Appl. Phys. Rev. **3**, 021101 (2016).

[3] H. Ryssel and I. Ruge, *Ionimplantation*, (B.G. Teubner, Stuttgart, 1978).

[4] V. Wang, J. W. Ward, and R. L. Seliger, J. Vac. Sci. Technol. **19**, 1158 (1981).

[5] T. Ishitani, K. Umemura, Y. Kawanami, and H. Tamura, J. Phys. Colloq. **45**, C9–191 (1984).

[6] P.D. Prewett and E.M. Kellogg, Nucl. Instrum. Methods Phys. Res. B **6**, 135 (1985).

[7] L. Bischoff, J. Teichert, S. Hausmann, T. Ganetsos, and G. Mair, Nucl. Instrum. Methods Phys. Res. B **161-163**, 1128 (2000).

[8] https://www.raith.com/products/velion.html?mobile=0

[9] ASM International, *Binary Alloy Phase Diagrams*, 2$^{nd}$ edition (1996).

[10] L. W. Swanson, A. E. Bell and G. A. Schwind, J. Vac. Sci. Technol. B **6**, 491 (1988).

[11] E. Hesse, L. Bischoff and J. Teichert, J. Phys. D: Appl. Phys. **27**, 427 (1994).

[12] W. M. Clark Jr., R. L. Seliger, M. Utlaut, A. E. Bell, and L.W. Swanson, J. Vac. Sci. Technol. B **5**, 197 (1987).

[13] H. Kamp and R. Schrepper, *Chemical formulae and data*, (Klett-Verlag, 1995, 75).

[14] W. Pilz, N. Klingner, L. Bischoff, P. Mazarov, and S. Bauerdick, J. Vac. Sci. Technol. B **37**, 021802-1 (2019).

[15] L. Bruchhaus, P. Mazarov, L. Bischoff, J. Gierak, A.D. Wieck, and H. Hövel, Appl. Phys. Rev. **4**, 011302 (2017).

[16] Ch. Akhmadaliev, L. Bischoff and B. Schmidt, Materials Science and Engineering C **26,** 818 (2006).